\documentclass[10pt,conference]{IEEEtran}
\IEEEoverridecommandlockouts
\usepackage{cite}
\usepackage{amsmath,amssymb,amsfonts}
\usepackage{algorithmic}
\usepackage{graphicx}
\usepackage{textcomp}
\usepackage{xcolor}

\usepackage{hyperref}
\hypersetup{
colorlinks = true, % false: boxed links; true: colored links
linkcolor=black, % color of internal links
citecolor=black, % color of links to bibliography
urlcolor=black % color of external links
}
\usepackage{amsmath,amssymb,amsfonts}
\usepackage{algorithmic}
\usepackage{graphicx}
\usepackage{textcomp}
\usepackage{xcolor}
\usepackage{xspace}
\usepackage{paralist}
\usepackage{amsthm}
\usepackage{booktabs}
\usepackage{braket}
\usepackage{qcircuit}
\usepackage{amssymb}% http://ctan.org/pkg/amssymb
\usepackage{pifont}% http://ctan.org/pkg/pifont
\newcommand{\cmark}{\ding{51}}%
\newcommand{\xmark}{\ding{55}}%

\theoremstyle{definition}
\newtheorem{definition}{\protect\definitionname}
\providecommand{\definitionname}{Definition}

\def\BibTeX{{\rm B\kern-.05em{\sc i\kern-.025em b}\kern-.08em
    T\kern-.1667em\lower.7ex\hbox{E}\kern-.125emX}}

\newcommand{\addSName}{\ensuremath{\mathtt{AS}}\xspace}
\newcommand{\bvName}{\ensuremath{\mathtt{BV}}\xspace}
\newcommand{\ceName}{\ensuremath{\mathtt{CE}}\xspace}

\newcommand{\qramName}{\ensuremath{\mathtt{QR}}\xspace}
\newcommand{\iqftName}{\ensuremath{\mathtt{IQ}}\xspace}

\newcommand{\qucat}{\texttt{QuCAT}\xspace}

\newcommand{\fun}[1]{\texttt{func#1}\xspace}
\newcommand{\uof}{\textit{uof}\xspace}
\newcommand{\wodf}{\textit{wodf}\xspace}

\newcommand{\testSuite}{\ensuremath{T}\xspace}
\newcommand{\quantumBitsSet}{\ensuremath{Q}\xspace}
\newcommand{\quantumProg}{\ensuremath{\textsc{QP}}\xspace}
\newcommand{\inputsSet}{\ensuremath{I}\xspace}
\newcommand{\outputsSet}{\ensuremath{O}\xspace}
\newcommand{\domain}[1]{\ensuremath{D_{#1}}\xspace}
\newcommand{\inputDom}{\ensuremath{\domain{I}}\xspace}
\newcommand{\outputDom}{\ensuremath{\domain{O}}\xspace}
\newcommand{\numRepIndex}{\ensuremath{n}\xspace}

\newcommand{\testInputValue}{\ensuremath{i}\xspace}

\newcommand{\progSpec}{\ensuremath{\mathtt{PS}}\xspace}

\def\BibTeX{{\rm B\kern-.05em{\sc i\kern-.025em b}\kern-.08em
    T\kern-.1667em\lower.7ex\hbox{E}\kern-.125emX}}
\begin{document}

\title{QuCAT: A Combinatorial Testing Tool for\\Quantum Software\thanks{This work is supported by the Qu-Test Project (No. 299827) funded by the Research Council of Norway. X. Wang is also supported by Simula's internal strategic project on quantum software engineering. P. Arcaini is supported by ERATO HASUO Metamathematics for Systems Design Project (No. JPMJER1603), JST, Funding Reference number 10.13039/501100009024 ERATO; and by Engineerable AI Techniques for Practical Applications of High-Quality Machine Learning-based Systems Project (Grant Number JPMJMI20B8), JST-Mirai. S. Ali acknowledges the support from the \textit{Quantum Hub initiative} (OsloMet).}
}

\author{\IEEEauthorblockN{Xinyi Wang}
\IEEEauthorblockA{
% \textit{dept. name of organization (of Aff.)} \\
\textit{Simula Research Laboratory}\\
Oslo, Norway \\
xinyi@simula.no}
\and
\IEEEauthorblockN{Paolo Arcaini}
\IEEEauthorblockA{
% \textit{dept. name of organization (of Aff.)} \\
\textit{National Institute of Informatics}\\
Tokyo, Japan \\
arcaini@nii.ac.jp}
\and
\IEEEauthorblockN{Tao Yue}
\IEEEauthorblockA{
% \textit{dept. name of organization (of Aff.)} \\
\textit{Simula Research Laboratory}\\
Oslo, Norway \\
tao@simula.no}
\and
\IEEEauthorblockN{Shaukat Ali}
\IEEEauthorblockA{
% \textit{dept. name of organization (of Aff.)} \\
\textit{Simula Research Laboratory and}\\
\textit{Oslo Metropolitan University}\\
Oslo, Norway \\
shaukat@simula.no}
% \and
% \and
}

\maketitle

\begin{abstract}
With the increased developments in quantum computing, the availability of systematic and automatic testing approaches for quantum programs is becoming increasingly essential. To this end, we present the quantum software testing tool \qucat for combinatorial testing of quantum programs. \qucat provides two functionalities of use. With the first functionality, the tool generates a test suite of a given strength (e.g., pair-wise). With the second functionality, it generates test suites with increasing strength until a failure is triggered or a maximum strength is reached. \qucat uses two test oracles to check the correctness of test outputs. We assess the cost and effectiveness of \qucat with 3 faulty versions of 5 quantum programs. Results show that combinatorial test suites with a low strength can find faults with limited cost, while a higher strength performs better to trigger some difficult faults with relatively higher cost.\\
\noindent Repository: \url{https://github.com/Simula-COMPLEX/qucat-tool}\\
Video: \url{https://youtu.be/UsqgOudKLio}
\end{abstract}


\section{Introduction}\label{sec:intro}
Quantum Software Engineering (QSE)~\cite{zhao2020quantum,SPP2022} is rapidly growing these years. QSE enables building software engineering solutions to solve complex problems with quantum software that can be executed on quantum computers. Testing quantum software is one area within QSE which presents some challenges due to the unique features of quantum computing, such as its probabilistic nature, destructive measurement, and no-cloning theorem of quantum states. Thus, it is essential to build automated solutions for testing quantum programs~\cite{MiranskyyICSE19,MiranskyyICSE2020,zhao2020quantum}. To this end, some approaches have been proposed, employing different techniques such as property-based testing~\cite{honarvar2020property}, search-based testing~\cite{genTestsQPSSBSE2021}, mutation testing~\cite{Fortunato2022,muttgGecco}, fuzz testing~\cite{quantfuzzposterICST2021}, metamorphic testing~\cite{AbreuQSE22}, in addition to input and output coverage criteria for quantum programs~\cite{ourICST2021}. These approaches are also supported by tools~\cite{Muskit,quitoASE21tool,qusbttool,FortunatoISSTA2022}.

%Combinatorial testing is a well-known software testing technique, which can be used to generate test suites to detect faults caused by interactions among parameters (e.g., pair-wise and 3-wise testing)~\cite{surveyCT2011}. It has been successfully applied in many domains (e.g., deep neural networks~\cite{Tao2019}, autonomous driving systems~\cite{Ma2019}). In a quantum program, a fault is often reflected as a faulty gate of one or more qubits in the quantum circuit corresponding to the quantum program. Test inputs are combinations of values initialized for qubits of the quantum circuit, aiming to trigger faults. Motivated by the success of combinatorial testing in testing classic software and analysis of causes of quantum faults, we proposed an approach in~\cite{CTQuantumQRS2021} for applying combinatorial testing for quantum programs. 
However, previous approaches overlook the types of inputs that can trigger a fault. Indeed, in a quantum program, a fault is often due to a faulty gate of the quantum circuit corresponding to the quantum program. Quantum gates combine the values of one or more qubits; therefore, a specific \emph{combination} of the values for some input qubits should be able to trigger the fault. For classic software, combinatorial testing is a well-known technique which is used to generate test suites to detect faults caused by specific combinations of input parameters (e.g., pair-wise or 3-wise testing)~\cite{surveyCT2011}. It has also been successfully applied in many other domains, such as deep neural networks~\cite{Tao2019} and autonomous driving systems~\cite{Ma2019}.

Motivated by the success of combinatorial testing in different domains, in~\cite{CTQuantumQRS2021} we proposed an approach for doing combinatorial testing of quantum programs.
However, the approach was not implemented in a tool usable by practitioners. To this aim, in this paper, we present how we engineered the approach in the tool \qucat. The tool can generate combinatorial test suites for quantum programs, execute them, and assess the test results automatically. Such quantum programs are written in the Qiskit framework~\cite{QiskitWille2019}. In Qiskit, a quantum circuit is a computational routine containing quantum wires, quantum gates, initialization, measurements, etc.

%We also apply PICT~\cite{pictTool} to generate combinatorial test suites. 

\qucat provides two functionalities of use. With functionality \fun{1}, \qucat can be used to generate a test suite with a specified \emph{strength}, while with functionality \fun{2}, \qucat can be used to generate test suites with incremental strengths until a fault is detected or the maximum strength is reached. \qucat employs two test oracles to assess test results. 
%To validate \qucat, we do experiments on 3 faulty versions of 5 different quantum programs to see cost and effectiveness of this tool.
We assess the cost and effectiveness of \qucat by experimenting it on 3 faulty versions of 5 different quantum programs.

%The rest of the paper is organized as follows: 
\noindent{\it Paper structure.}
Sect.~\ref{sec:background} briefly introduces qubits and quantum circuits and provides the definitions of key concepts. We describe the overview and methodology of \qucat in Sect.~\ref{sec:methodology}. Validation of the approach is presented in Sect.~\ref{sec:validation}. Finally, we conclude our work and discuss the future work in Sect.~\ref{sec:conclusion}.

\section{Preliminaries}\label{sec:background}

\subsection{Qubits and Quantum Circuits}
Classic computers work on classic bits, which can be either 0 or 1 assigned to each bit. Differently, a quantum computer works on \emph{qubits}, which can be both 0 and 1 simultaneously, i.e., they are in \emph{superposition} until being observed.

\begin{figure}[!t]
\footnotesize
\centerline{
\Qcircuit @C=1em @R=1em {
\lstick{q_{0} \ket{0}} & \gate{H} &\ctrl{1} & \meter & \qw \\
\lstick{q_{1} \ket{0}}  & \qw & \targ & \meter & \qw\\
}
}
\caption{Entanglement Quantum Circuit}
\label{fig:entanglement}
\end{figure}
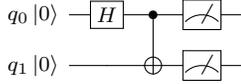

For a qubit, we define its quantum state $\ket{\psi}$ as follows:
\begin{equation*}
    \ket{\psi}=\alpha\ket{0}+\beta\ket{1}
\end{equation*}
where $\alpha$ and $\beta$ are two complex numbers, each defining the \emph{amplitude} of the qubit in terms of \emph{magnitude} and \emph{phase}. $|\alpha|^{2}$ and $|\beta|^{2}$ show the probability for a qubit to be in state $\ket{0}$ or $\ket{1}$, and it holds that $|\alpha|^{2}+|\beta|^{2}=1$. An example of quantum circuits is shown in Fig.~\ref{fig:entanglement}, which is an Entanglement program.
First, we can initialize the two qubits to a state. For instance, in this example, we initialize both of them as the $\ket{0}$ state. Then, we apply a Hadamard gate on the first qubit (i.e., $q_0$) to turn it into superposition and a CX gate on both qubits, i.e., $q_0$ as the control qubit and $q_1$ as the target qubit. If $q_0$ is in state $\ket{1}$, $q_1$ will be flipped. Otherwise, $q_1$ won't be flipped. Finally, we measure the two qubits to obtain the outputs. For this circuit, there will be a 50\% probability of obtaining output 00 and a 50\% probability of obtaining 11.

\subsection{Definitions}\label{sec:definition}
This section provides definitions of key concepts of quantum programs and combinatorial testing.

%including the method to generate test suites with combinatorial testing.

\begin{definition} [Inputs, outputs, and quantum program]\label{def:basicquantum}
Let $\quantumBitsSet=$ $\{q_1,$ $q_2, \ldots, q_n\}$ be the set of qubits of a quantum program \quantumProg. A subset $\inputsSet \subseteq \quantumBitsSet$ are the \emph{input qubits}, while a subset $\outputsSet \subseteq \quantumBitsSet$ are the \emph{output qubits}. $\inputDom = \mathcal{B}^{|\inputsSet|}$ and $\outputDom = \mathcal{B}^{|\outputsSet|}$ are the input and output values. A quantum program can be described as a function $\quantumProg \colon \inputDom \rightarrow 2^{\outputDom}$.
\end{definition}

According to the definition, a quantum program can return different outputs with the same input value due to its probabilistic nature. If the expected behaviour of the quantum program is known, we can define it as a program specification.

\begin{definition} [{Program specification}]\label{def:programspecification}
Given a quantum program \quantumProg, \progSpec is identified as the \emph{program specification} which is the expected behaviour of \quantumProg. Given an input $\testInputValue \in \inputDom$ and a possible output value $h$, \progSpec states the expected probability of occurrence of output $h$ for input $i$ (i.e., $\progSpec(i, h)=p_h$).
\end{definition}

%When testing a quantum program, to compare with the program specification with the probabilities of occurrences of outputs of an input, the quantum program has to be executed for multiple times to get the distributions of output values due to the non-deterministic nature of quantum programs.
Due to its non-deterministic nature, for a given input, a quantum program must be executed multiple times to get the distribution of output values.

\begin{definition} [Test inputs and test results]\label{def:testinputresult}
A \emph{test input} is defined by the pair $\langle \testInputValue, \numRepIndex \rangle$, where \testInputValue is the assignment to input qubits \inputsSet (i.e., $\testInputValue \in \inputDom$), and \numRepIndex represents the number of times that \quantumProg needs to be executed with \testInputValue. \emph{Test results} are defined as $res = [\quantumProg(i), \ldots, \quantumProg(i)] = [o_1, \ldots, o_{\numRepIndex}]$, where $o_j$ is the output returned by \quantumProg at the $j$th execution.

%For one input, we need to execute multiple times to get the distribution of outputs.
We define two failure types to assess the test results.
\begin{itemize}
    \item \emph{Unexpected Output Failure} (\uof): for a given input \testInputValue, it checks whether there exists one output $o_j$ that is unexpected (i.e., $\progSpec(i, o_j)=0$).
    \item \emph{Wrong Output Distribution Failure} (\wodf): when there is no unexpected output, it checks whether there is a significant difference between the observed distribution of outputs and that defined in the program specification \progSpec.
    %, which should be assessed with a statistical test.
    We use \emph{goodness of fit test} with Pearson's chi-square test~\cite{agresti2019introduction} to assess the difference.
\end{itemize}
\end{definition}

In a quantum program, faults may be caused by the incorrect use of gates, and they could be exposed by some specific values combination of some of the inputs. Combinatorial testing can be applied to find such combinations for testing quantum programs. Value schema~\cite{surveyCT2011} specifies a combination of input values that has to be covered in a test.

\begin{definition} [{Value schema}]\label{def:valueschema}
For a quantum program with input qubits $(q_0, q_2, \ldots, q_{|\inputsSet|-1})$, a \emph{k-value schema} ($k > 0$ and $k \le |\inputsSet|$) is $(\ldots, v_{i_1}, \ldots, v_{i_k}, \ldots)$, where $k$ qubits are assigned with fixed values while the others are not fixed.

Here, we identify with '-' the values of qubits, which are not fixed. For example, given a quantum program with 4 input qubits in which the second and third qubits are fixed with '0' and '1', the \emph{2-value schema} is $(-, 0, 1, -)$.
\end{definition}

\begin{definition} [Combinatorial test suite]\label{def:combinatorialtestsuite}
%Each \emph{k-value schema} is contained in a \emph{combinatorial test suite} $T$ with strength $k$ which shows types of testing (e.g., pairwise testing when $k = 2$). For a quantum program, each \emph{k-value schema} in at least one input value of $T_k$.
A \emph{combinatorial test suite} $T_k$ of strength $k$ contains, for each \emph{k-value schema}, at least a test covering the schema.
\end{definition}

\section{QuCAT Tool Description and Methodology}\label{sec:methodology}
\begin{figure}[!tb]
\centering
\includegraphics[width=0.85\columnwidth]{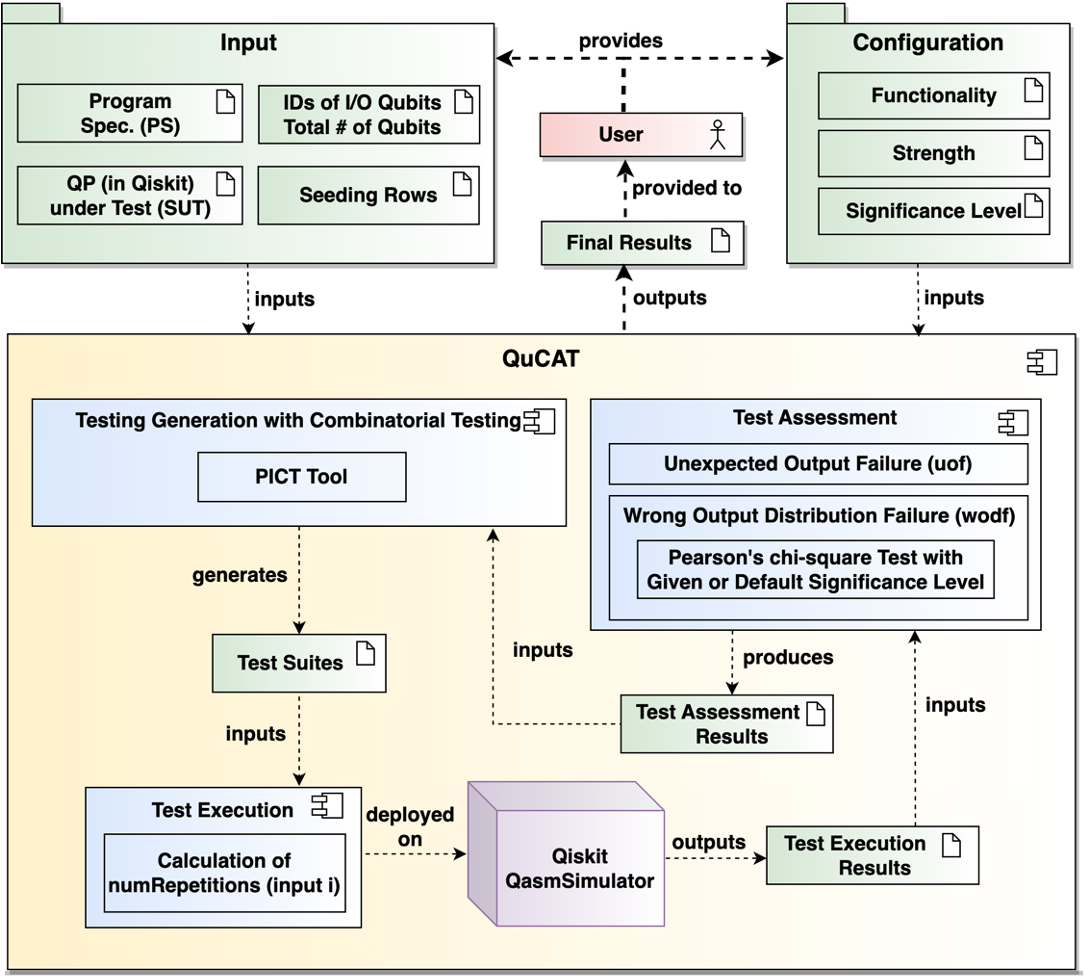}
\caption{Architecture of the \qucat Tool}
\label{fig:overview}
\end{figure}

We present how we have engineered the approach proposed in~\cite{CTQuantumQRS2021} in the tool \qucat, which automatically generates combinatorial test suites, executes them, and analyzes test results. The overview of \qucat is shown in Fig.~\ref{fig:overview}.

\subsection{Functionalities}
\qucat can be used in two ways:
\begin{itemize}
    \item \fun{1}: A user specifies a fixed strength $k$ and the tool generates and assesses a combinatorial test suite $T_k$.
    \item \fun{2}: Test suites are kept on being generated with an incremental strength, beginning from $k = 2$, until a fault is found or the maximum strength $K$ is reached.
\end{itemize}

\subsection{Input and Configuration}\label{sec:configure}
To use \qucat, a user should provide \emph{input} information and the tool \emph{configuration}. The \emph{input} information includes the quantum program under test (SUT), the program specification (PS), the list of input and output qubits, and the total number of qubits. The \emph{configuration} contains parameters of \qucat:
\begin{itemize}
\item the functionality to use \qucat, i.e., \fun{1} or \fun{2} as described in Sect.~\ref{sec:definition}.
\item a \emph{strength} value. If the user selects \fun{1}, the strength is the value $k$ for the combinatorial test suite $\testSuite_k$ to generate (e.g., $k=2$ for pairwise testing, $k=3$ for 3-wise testing). In this case, the default value is $k=2$. If instead, the user selects functionality \fun{2}, the value is the maximum strength $K$ until which the tool must generate test suites.
\item \qucat employs Pearson's chi-square test to assess failure type \wodf (see Sect.~\ref{sec:definition}). The user can specify the significance level for the test. The default value is $0.01$.
\item Moreover, the user can provide a file of \emph{seeding rows}, which contains test cases required to appear in all generated test suites by combinatorial testing. 
\end{itemize}
Then, \qucat generates tests following the configuration file.
%In particular, if the user chooses $w_2$, strength $k$ refers to the maximum strength (i.e., \qucat keeps on generating test suites until the test finds one fault or the strength reaches $K$).

%When using the tool, the user can first select the functionalities described in Sect.~\ref{sec:definition} (i.e., \fun{1} or \fun{2}). Moreover, if needed, the user can provide a file of \emph{seeding rows} which contains test cases that the user requires to appear in all generated test suites by combinatorial testing. Then, \qucat begins to generate test suites according to the configuration file. In particular, if the user chooses $w_2$, strength $k$ refers to the maximum strength (i.e., \qucat keeps on generating test suites until the test finds one fault or the strength reaches $K$).

In summary, as shown in Fig.~\ref{fig:overview}, when using \qucat, the user should
\begin{inparaenum}[(i)]
    \item select one of the two functionalities,
    \item provide input information of the quantum program,
    \item configure parameters for combinatorial testing and test assessment, and
    \item provide a file of seeding rows if needed.
\end{inparaenum}
%
%Specifically, configuration parameters have one default setting and the user can specify them according to requirements, together with input information in one configuration file. Also, the file of seeding rows is not mandatory.

\subsection{Process of Test Generation, Execution, and Assessment}\label{sec:process}
We describe the steps performed by \qucat to generate test suites with combinatorial testing, execute them over the quantum program, get test results, and perform test assessment. 
\begin{itemize}
%\item \qucat employs PICT~\cite{pictTool} to generate combinations of input values of input qubits. Each qubit will be assigned by one value for each input \testInputValue, which represents one test case.
\item \qucat employs PICT~\cite{pictTool} to generate combinatorial test suites of a given strength (see Def.~\ref{def:combinatorialtestsuite}). Dependent on the selected functionality, \qucat works as follows:
\begin{itemize}
    \item If the user chooses \fun{1}, the tool generates a test suite $T_k$ for the strength $k$ defined by the user.
    \item If the user chooses \fun{2}, the tool keeps on generating combinatorial test suites with incremental strengths until a test triggers a failure or the maximum strength value $K$ is reached. Note that \qucat will assess results every time when a new test suite is generated.
\end{itemize}
\item For a test input $\langle \testInputValue, \numRepIndex \rangle$, the number of executions \numRepIndex is defined according to the program specification.
%as the number of required repetitions is proportional to the number of possible values.
In detail, the number of repetitions is computed as the number of possible outputs for input $\testInputValue$ multiplied by 100 (i.e., $\numRepIndex = |\{h \in \outputDom \mid \progSpec(i,h) \neq 0\}| \times 100$).
\item \qucat automatically checks each test case (i.e., test input) according to two failure types \uof and \wodf, as defined in Def.~\ref{def:testinputresult}, by comparing test results with the program specification.
\begin{itemize}
    \item \qucat first checks failure type \uof. For input \testInputValue, after \numRepIndex executions, if there is any unexpected output value that does not appear in the program specification, the test case fails for \uof.
    \item If there is no unexpected output, \qucat checks \wodf. After \numRepIndex executions, if the output distribution is significantly different from that of the program specification (checked with the statistical test introduced in Def.~\ref{def:testinputresult}), the test fails for \wodf.
\end{itemize}
\item After the generation of all the test suites, their execution, and their assessment, \qucat provides the following files as output:
\begin{inparaenum}[(i)]
    %\item a list of input values test results and assessment results for each test case.
    \item a file containing execution results of all test cases;
    \item a file of assessments of each test case;
    \item unit tests written in the \texttt{unittest} framework for the generated test suites that the user can use to rerun the test cases when debugging the quantum program.
\end{inparaenum}
\end{itemize}
% %
% \begin{table}[!tb]
% \centering
% \caption{Data for \usageScenario{1}}
% \label{table:expResultsRQ1RQ2}
% % \setlength{\tabcolsep}{2pt}
% \resizebox{\columnwidth}{!}{
% \begin{tabular}{cccc|ccc|ccc}
% \toprule
%  & \multicolumn{3}{c}{$k=2$} & \multicolumn{3}{c}{$k=3$} & \multicolumn{3}{c}{$k=4$}\\
% \cline{2-4}\cline{5-7}\cline{8-10}\\
%  & F$_{1}$ & F$_{2}$ & F$_{3}$ & F$_{1}$ & F$_{2}$ & F$_{3}$ & F$_{1}$ & F$_{2}$ & F$_{3}$ \\
% \midrule
% \addSName & \cmark & \cmark & \xmark & \cmark & \cmark & \xmark & \cmark & \cmark & \cmark\\
% \bottomrule
% \end{tabular}
% }
% \end{table}
%
\begin{table*}[!tb]
\centering
%\small
\caption{Results for functionality \fun{1}}
\label{table:usagescenario1}
\setlength{\tabcolsep}{4.5pt}
% \resizebox{\columnwidth}{!}{
\begin{tabular}{cccccccc|cccccc|cccccc}
\toprule
&  & \multicolumn{6}{c}{$k=2$} & \multicolumn{6}{c}{$k=3$} & \multicolumn{6}{c}{$k=4$}\\
\cline{3-8}\cline{9-14}\cline{15-20}\\ 
QP& $|\inputsSet|$ & \multicolumn{2}{c}{F$_1$} & \multicolumn{2}{c}{F$_2$} & \multicolumn{2}{c}{F$_3$} & \multicolumn{2}{c}{F$_1$} & \multicolumn{2}{c}{F$_2$} & \multicolumn{2}{c}{F$_3$} & \multicolumn{2}{c}{F$_1$} & \multicolumn{2}{c}{F$_2$} & \multicolumn{2}{c}{F$_3$}\\
& & failure & $|T_2|$ & failure & $|T_2|$ & failure & $|T_2|$ & failure & $|T_3|$ & failure & $|T_3|$ & failure & $|T_3|$ & failure & $|T_4|$ &failure & $|T_4|$ & failure & $|T_4|$ \\
\midrule
\addSName& 6 & \cmark & 6 & \cmark & 6 & \xmark & 6 & \cmark & 14 & \cmark & 13 & \xmark & 13 & \cmark & 26 & \cmark & 27 & \cmark & 24\\
\bvName&7 & \cmark & 7 & \xmark & 6 & \xmark & 7 & \cmark & 15 & \cmark & 14 & \cmark & 14 & \cmark & 31 & \cmark & 30 & \cmark & 32\\
\ceName&11 & \cmark & 8 & \xmark & 9 & \xmark & 8 & \cmark & 19 & \cmark & 19 & \cmark & 19 & \cmark & 47 & \cmark & 43 & \cmark & 46\\
\iqftName&10 & \cmark & 8 & \cmark & 9 & \xmark & 8 & \cmark & 18 & \cmark & 18 & \cmark & 18 & \cmark & 43 & \cmark & 44 & \cmark & 44\\
\qramName&9 & \cmark & 8 & \cmark & 8 & \xmark & 8 & \cmark & 14 & \cmark & 17 & \xmark & 16 & \cmark & 38 & \cmark & 38 & \cmark & 38\\
\bottomrule
\end{tabular}
% }
\end{table*}

\section{Tool Validation}\label{sec:validation}

We evaluate \qucat with five quantum programs. For each program, we created three faulty versions (i.e., F$_1$, F$_2$ and F$_3$). F$_3$ is the most difficult one, followed by F$_2$ and F$_1$ based on the number of input values that can trigger the fault. The selected programs are: 1) \bvName: the Bernstein-Vazirani cryptography algorithm; 2)  \qramName: quantum random access memory implementation; 3) \iqftName: inverse quantum Fourier transform; 4) \addSName: performing $v1 = v1 + v2 * v2$, where $v1$ and $v2$ are two quantum integers; and 5) \ceName: a conditional addition on a quantum integer. The characteristics of these quantum programs vary in terms of the number of qubits and gates and circuit depth. The gate numbers \bvName, \qramName, \iqftName, \addSName, \ceName are 21, 15, 60, 25, and 25, and the depths are 3, 12, 56, 22, and 26.

% \begin{inparaenum}[(i)]
%\begin{itemize}
%\item \bvName: a cryptography program, the Bernstein-Vazirani cryptographic algorithm;
%\item \qramName: an algorithm to manage quantum random access memory and performing mathematical operation;
%\item \iqftName: inverse quantum Fourier transform; 
%\item \addSName: a mathematical operation of a quantum integer and a squared quantum integer;
%\item \ceName: a conditional addition on an quantum integer.
%\end{itemize}
% \end{inparaenum}

Table~\ref{table:usagescenario1} shows the results of testing the faulty versions of the five quantum programs with functionality \fun{1} using values 2, 3, and 4 for strength $k$. $|\inputsSet|$ is the number of input qubits and $|T_k|$ is the test suite size. Column \emph{failure} tells whether a failure is triggered by the test suite (i.e., a fault is detected). Higher numbers of qubits and strength $k$ (e.g., $k=4$) result in a higher test suite size, i.e., the higher cost in terms of test suite size (e.g., \ceName and \iqftName). One observation is that even with the lower strength (i.e., pairwise testing), \qucat can still find the easiest faults (i.e., F$_1$) in all the programs. All the faults, including the most difficult one (i.e., F$_3$), are triggered by the test suites generated with strength 4.

%Table~\ref{table:usagescenario1} shows results of testing quantum programs under \usageScenario{1}. $|\inputsSet|$ is the number of input qubits and $|T_k|$ is test suite size. The higher number of qubits and strength $K$ (e.g., $k=4$) result in a higher test suite size For example, \ceName and \iqftName contributes to higher cost as expected. In addition, even if cost for lower strength (e.g., pairwise testing) is low, it can still find faults of all programs with fault number 1 (i.e., the easiest fault to be detected). All faults, even the most difficult ones (i.e., F$_3$) can be triggered by test suites with strength 4.

Table~\ref{table:usagescenario2} reports the results of running \qucat with functionality \fun{2} using 4 as maximum strength $K$.
\begin{table}[!tb]
\centering
%\small
\caption{Results for functionality \fun{2}}
\label{table:usagescenario2}
% \setlength{\tabcolsep}{2pt}
% \resizebox{\columnwidth}{!}{
\begin{tabular}{ccc|cc|cc}
\toprule
& \multicolumn{2}{c}{F$_1$} & \multicolumn{2}{c}{F$_2$} & \multicolumn{2}{c}{F$_3$}\\
\cline{2-3}\cline{4-5}\cline{6-7}\\ 
% QP & \multicolumn{2}{c}{F$_1$} & \multicolumn{2}{c}{F$_2$} & \multicolumn{2}{c}{F$_3$}\\
QP & $|T^c|$ & $k_{\mathit{end}}$ & $|T^c|$ & $k_{\mathit{end}}$ & $|T^c|$ & $k_{\mathit{end}}$\\
\midrule
\addSName& 5 & 2 & 8 & 3 & 32 & 4\\
\bvName& 4 & 2 & 6 & 2 & 39 & 4\\
\ceName& 2 & 2 & 6 & 2 & 13 & 3\\
\iqftName& 1 & 2 & 15 & 3 & 25 & 3\\
\qramName& 2 & 2 & 6 & 2 & 26 & 4\\
\bottomrule
\end{tabular}
% }
\end{table}
%
%$|T^c|$ reports the number of executed tests before detecting the fault, and $k_{\mathit{end}}$ refers to the strength of the generated test suite containing the test case which triggered the fault.
$k_{\mathit{end}}$ refers to the strength of the generated test suite containing the test case which triggered the fault; $|T^c|$ reports the number of executed tests, i.e., all the tests in the generated test suites $T_{2}$, \ldots, $T_{k_{\mathit{end}} - 1}$ and the tests in $T_{k_{\mathit{end}}}$ that have been executed before triggering the fault (including the failing test).
F$_1$ faults for all the programs and F$_2$ faults of three programs are triggered by test suites with strength 2. Test suites with strength 3 have been necessary to detect faults of two programs with F$_2$ faults and two programs with F$_3$ faults. For the rest of F$_3$ faulty programs, \qucat had to generate till the test suites with the maximum strength 4, which were able to detect these faults.

\section{Conclusions and Future Work}\label{sec:conclusion}
We presented \qucat-- an automated tool for combinatorial testing of quantum programs. \qucat uses two test oracles to assess test results and two functionalities to generate tests. We also presented the architecture of the tool. Finally, we validated the tool by testing three faulty versions of five quantum programs to assess the cost and effectiveness of \qucat followed by discussing the results. In the future, we plan to extend the tool further to provide features such as pinpointing the location where the program failed (i.e., fault localisation) and porting the tool for real quantum computers.

\bibliographystyle{IEEEtran}
\bibliography{qucat_ASE2023demo}

\end{document}